\def\tit#1{``#1,''}
\long\def\comment#1{}

\let\ox=\otimes

\newcommand{\Mx}[3]{\left#1\begin{array}{rrrr}#2\end{array}\right#3}

\newcommand{\ket}[1]{| #1 \rangle}
\newcommand{\bra}[1]{\langle #1 |}

\newcommand{\vxv}[1]{\ket{#1}\bra{#1}}
\newcommand{\brkt}[2]{\langle #1 \mid #2 \rangle}

\newcommand\Hil{\mathcal H}
\newcommand\C{\mathbb C}
\newcommand\R{\mathbb R}

\newcommand\QPU{\mathsf{Q}_{\mathsf{PU}}}

\newcommand\Shft{\mathsf {S_{hft}}}

\newcommand{\wave}[1]{\hbox to #1{\leaders\hbox{${\sim}\!$}\hfil}}
\newcommand{\rightwave}{\mathop{\wave{2cm}\!{\to}}\limits}

\newcommand{\GE}{\geqslant}

\newcommand{\op}[1]{\boldsymbol{#1}}

\newcommand\eq[1]{Eq.~(\ref{#1})}

\documentclass[12pt]{article}
\usepackage{exscale}
\usepackage{amssymb,amsbsy}

\begin{document}
\title{\bf Von Neumann Quantum Processors\thanks{
304th WE-Heraeus-Seminar `Elementary Quantum-Processors,' 13--15 October 2003}}
\author{{\em Alexander Yu.\ Vlasov%
 \thanks{Electronic mail: \tt Alexander.Vlasov@PObox.spbu.ru}}\\
 Federal Radiological Center (IRH)\\
 197101 Mira Street 8, St.--Petersburg, Russia}
\date{}
%\sloppy
\maketitle
\renewcommand{\abstractname}{{\large ABSTRACT}}
\begin{abstract}
\noindent
Most modern classical processors support so-called {\em von Neumann
architecture} with program and data registers. In present work is revisited
similar approach to models of quantum processors. Deterministic programmable
quantum gate arrays are considered as an example. They are also called
{\em von Neumann quantum processors} here and use conditional quantum dynamics.
Such devices have some problems with universality, but consideration
of hybrid quantum processors, {\em i.e.}, models with both continuous and
discrete quantum variables resolves the problems. It is also discussed
comparison of such a model of quantum processors with more traditional
approach.
\end{abstract}

%\newpage
\section{Introduction}
\label{sec:intro}
So-called von Neumann architecture was used already in first electronic
computers like {\sf EDVAC} \cite{EDVAC} more than 50 years ago.
Main idea of such approach is storage of program in computer memory, unlike
of some elementary electronic calculators.

Such idea is also appropriate for quantum computers and present work
recollects some previous discussions \cite{Vla01a,Vla01b,Vla02,Vla03}.
General idea of such aproach --- is to consider a Hilbert space of composite
quantum system with two parts:
\begin{equation}
 \Hil = \Hil_c \ox \Hil_d.
\label{H12}
\end{equation}
Here $\Hil_d$ is a Hilbert space of quantum system considered as {\em data},
and $\Hil_c$ is Hilbert space of {\em ``code''}, i.e., {\em program}.

\section{Quantum processing units}

\subsection{Programmable quantum gate arrays}

The programmable quantum gate array \cite{Vla01a,NC97,VC00,VMC01} or
{\em quantum processor} \cite{Vla01b,HBZ01} --- is a quantum
circuit with fixed structure.
Similarly with classical von Neumann Architecture here are
{\em data register} $\ket{\psi} \in \Hil_d$ and {\em program (code) register}
$\ket{c} \in \Hil_c$. Different operations $\op u$
with data are indexed by a state of the program (code)
\begin{equation}
\QPU \colon \bigl(\ket{c} \ox \ket{\psi}\bigr)
    \mapsto \ket{c'} \ox (\op u_c\ket{\psi}).
\label{qproc}
\end{equation}

It was discussed already in \cite{NC97}, that \eq{qproc} is compatible with
unitary quantum evolution, if different states of {\em program register}
are orthogonal --- due to such requirement number of accessible programs
coincides with dimension of Hilbert space and it produces some challenge for
construction of universal quantum processors. It was suggested few ways
around such a problem: to use specific ``stochastic'' design of universal
quantum processor \cite{NC97,VC00,VMC01,HBZ01},
to construct (non-stochastic) quantum processor with possibility to
approximate any gate with given precision \cite{Vla01a,Vla01b,VC00,VMC01}
(it is also traditional approach to universality \cite{Deu85,Deu89,Ek95},
sometime called ``universality in approximate sense'' \cite{Cle99}).

\subsection{Conditional quantum dynamics}

In finite-dimensional case unitary operator $\QPU$ satisfying \eq{qproc}
can be simply found \cite{Vla01a,Vla01b}. Let us consider case with
$\ket{c'} = \ket{c}$ in \eq{qproc}. It was already mentioned,
that states $\ket{c}$ of program register corresponding to different
operators $\op u_c$ are orthogonal and, so, may be chosen as basis. In such
a basis $\op u_c$ is simply set of matrices numbered by integer index $c$,
and operator $\QPU$ \eq{qproc} can be written as block-diagonal
$NM \times NM$ matrix:
\begin{equation}
\QPU = \Mx({\op u_1\\&\op u_2&&\smash{\mbox{\Huge$0$}}\\
                 &&\ddots\\\smash{\mbox{\Huge$0$}}&&&\op u_M}),
\end{equation}
with $N \times N$ matrices $\op u_c$,
if dimensions of program and data registers are $M$ and $N$ respectively;
\begin{equation}
\QPU = \sum_{c=1}^M \ket{c}\bra{c} \ox \op u_c,
\label{conddyn}
\end{equation}
It is {\em conditional quantum dynamics} \cite{Joz95}.
For quantum computations with qubits $M=2^m$, $N=2^n$.

\subsection{Three-buses design}

Operator $\QPU$ described above is only first approach to von Neumann
quantum processors. Such {\em quantum processing unit} (QPU) simply generates
some gate en{\em coded} (indexed) by state of program register. For
performing arbitrary sequence of operations it is possible to use more
difficult design \cite{Vla01a,Vla01b,Vla03}.

\begin{figure}[t]
\begin{center}
\unitlength=1mm
\begin{picture}(140.00,85)(3,0)
\thinlines
\put(5,65.00){\dashbox{2}(50,20)[cc]{Pseudo-classical bus ($\Hil_p$)}}
\put(5,35.00){\dashbox{2.00}(50,20.00)[cc]{Intermediate bus ($\Hil_c$)}}
\put(5.00,5.00){\dashbox{2}(50.00,20.00)[cc]{Quantum data bus ($\Hil_d$)}}
\thicklines
\put(55.00,10.00){\line(1,0){45}}
\put(55.00,11.00){\line(1,0){45}}
\put(55.00,12.00){\line(1,0){45}}
\put(125.00,10.00){\line(1,0){15}}
\put(125.00,11.00){\line(1,0){15}}
\put(125.00,12.00){\line(1,0){15}}
\put(55.00,18.00){\line(1,0){45.00}}
\put(55.00,19.00){\line(1,0){45.00}}
\put(55.00,20.00){\line(1,0){45.00}}
\put(125.00,18.00){\line(1,0){15.00}}
\put(125.00,19.00){\line(1,0){15.00}}
\put(125.00,20.00){\line(1,0){15.00}}
\put(55,40.00){\line(1,0){10.00}}
\put(55,41.00){\line(1,0){10.00}}
\put(55,42.00){\line(1,0){10.00}}
\put(90,40.00){\line(1,0){10}}
\put(90,41.00){\line(1,0){10}}
\put(90,42.00){\line(1,0){10}}
\put(125.00,40.00){\line(1,0){15}}
\put(125.00,41.00){\line(1,0){15}}
\put(125.00,42.00){\line(1,0){15}}
\put(55,48.00){\line(1,0){10}}
\put(55,49.00){\line(1,0){10}}
\put(55,50.00){\line(1,0){10}}
\put(90.00,48.00){\line(1,0){10.00}}
\put(90.00,49.00){\line(1,0){10.00}}
\put(90.00,50.00){\line(1,0){10.00}}
\put(125.00,48.00){\line(1,0){15.00}}
\put(125.00,49.00){\line(1,0){15.00}}
\put(125.00,50.00){\line(1,0){15.00}}
\put(55.00,70.00){\line(1,0){10}}
\put(55.00,71.00){\line(1,0){10}}
\put(55.00,72.00){\line(1,0){10}}
\put(90.00,70.00){\line(1,0){50}}
\put(90.00,71.00){\line(1,0){50}}
\put(90.00,72.00){\line(1,0){50}}
\put(55.00,78.00){\line(1,0){10}}
\put(55.00,79.00){\line(1,0){10}}
\put(55.00,80.00){\line(1,0){10}}
\put(90,78.00){\line(1,0){50}}
\put(90,79.00){\line(1,0){50}}
\put(90,80.00){\line(1,0){50}}
\put(65,35.00){\framebox(25,50.00)[cc]{
 \shortstack{Reversible\\program\\($\Shft$)}}}
\put(100,5.00){\framebox(25,50.00)[cc]{
 \shortstack{Quantum\\processing\\unit ($\QPU$)}}}
\put(60,75.00){\makebox(0,0)[cc]{$\cdots$}}
\put(60,45.00){\makebox(0,0)[cc]{$\cdots$}}
\put(77.50,15.00){\makebox(0,0)[cc]{$\cdots$}}
\put(112.50,75.00){\makebox(0,0)[cc]{$\cdots$}}
\put(132.50,45.00){\makebox(0,0)[cc]{$\cdots$}}
\put(132.50,15.00){\makebox(0,0)[cc]{$\cdots$}}
\end{picture}
\end{center}

\caption{Von Neumann quantum processor with three buses.}
\label{Fig3Buses}
\end{figure}
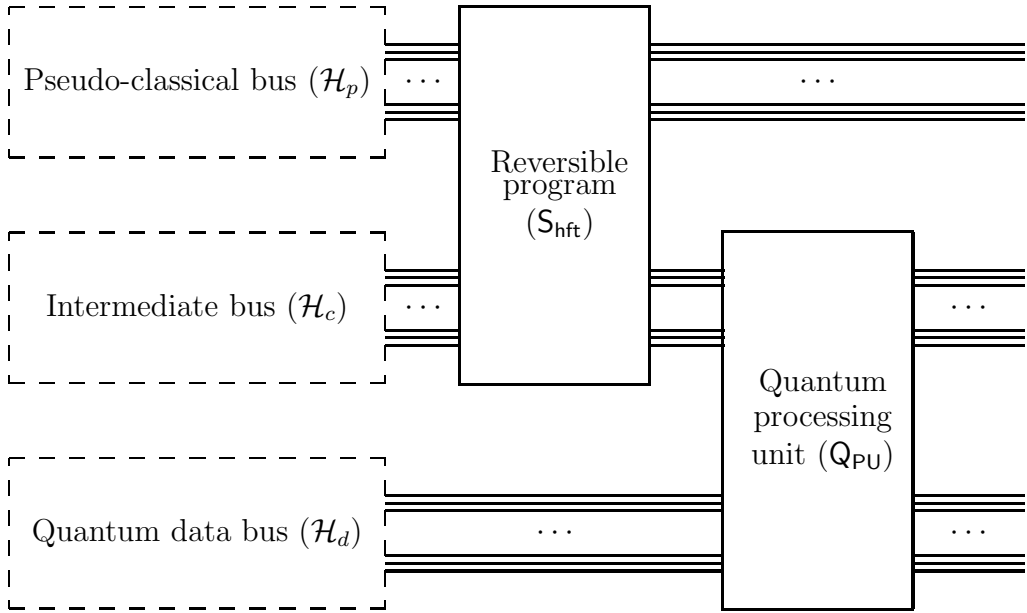

Let us instead of one program (code) register $\Hil_c$ consider
compound system $\Hil_p \ox \Hil_c$, there $\Hil_p$ corresponds to
``program memory'' in von Neumann architecture. Then Hilbert space
of such von Neumann quantum processor may be described as
\begin{equation}
 \Hil = \Hil_p \ox \Hil_c \ox \Hil_d.
\label{H123}
\end{equation}

Let us use simplest example with ``cyclic memory'' (resembling
{\em Hg delayed-line memory} of some old classical computers).
In this example $\dim \Hil_c = M$ and it is necessary to have
possibility to perform any sequence of up to $P$ operators $\op u_c$
with data register. Then it is possible to consider $\dim \Hil_p = M^{P}$
(it is $mP$ qubits for $M=2^m$) and
write basic states $\ket{p}\ket{c} \in \Hil_p \ox \Hil_c$ as
\begin{equation}
 \ket{p}\ket{c} \equiv \ket{p;c} \equiv \ket{c_P,\ldots,c_1;c_0},
\end{equation}
where $c_i < M$ are integer indexes and $c_0 \equiv c$ is index in $\Hil_c$.
It is possible to consider {\em right cyclic shift operator} $\Shft$
acting on $\Hil_p \ox \Hil_c$ as
\begin{equation}
 \Shft \colon \ket{c_P,c_{P-1}\ldots,c_1;c_0}
  \mapsto \ket{c_0,c_P,\ldots,c_2;c_1}
\label{Shft}
\end{equation}

Now it is possible to compose application of $\Shft$ \eq{Shft} operator on
first two terms in \eq{H123} \{$\Hil_p \ox \Hil_c$\} with $\QPU$ on
$\Hil_c \ox \Hil_d$. Then it is possible to write for
$P$ applications of such compound operator $\Shft\QPU$
\begin{equation}
 (\Shft\QPU)^P \colon \bigl(\ket{p;c}\ket{\psi}\bigr) \mapsto
 \ket{p;c}(\op u_{c_{P}}\cdots\op u_{c_1}\ket{\psi})
\label{CtrlShft}
\end{equation}
and it is possible to implement any sequence with $P$ operators using
different programs $\ket{p;c}$.

It should be mentioned, that instead of cyclic shift operator it is possible
to use arbitrary reversible algorithm for generation of indexes in
register $\ket{c}$ and so general three-buses design may be depicted
on Fig.~\ref{Fig3Buses}. Here Hilbert spaces $\Hil_p$, $\Hil_c$ and
$\Hil_d$ are called {\em pseudo-classical, intermediate and quantum data
buses} respectively \cite{Vla01a,Vla01b,Vla03}.

\section{Hybrid quantum processors}\label{sec:hybproc}

\begin{figure}[ht]
\bigskip
\begin{center}
\unitlength=1.2mm
\linethickness{0.4pt}
\begin{picture}(90,75)
\put(10,75){\makebox(0,0)[lc]{\wave{70\unitlength}}}
\put(10,70){\makebox(0,0)[lc]{\wave{70\unitlength}}}
%{\line(1,0){70}}
\put(10,65){\makebox(0,0)[lc]{\wave{70\unitlength}}}
%{\line(1,0){70}}
\put(10,60){\line(1,0){70}}
\put(10,55){\makebox(0,0)[lc]{\wave{70\unitlength}}}
%{\line(1,0){70}}
\put(10,50){\makebox(0,0)[lc]{\wave{70\unitlength}}}
%{\line(1,0){70}}
\put(10,45){\makebox(0,0)[lc]{\wave{70\unitlength}}}
\put(10,40){\line(1,0){70}}
\put(10,30){\line(1,0){70}}
\put(10,20){\line(1,0){70}}
\put(10,10){\line(1,0){70}}
\put(15,75){\vector(0,-1){45}}
\put(15,75){\circle*{1.5}}
\put(20,70){\vector(0,-1){40}}
\put(20,70){\circle*{1.5}}
\put(25,65){\circle*{1.5}}
\put(25,65){\vector(0,-1){35}}
\put(30,60){\vector(0,-1){40}}
\put(30,60){\circle*{1}}
\put(30,30){\circle*{1}}
\put(35,55){\vector(0,-1){35}}
\put(35,55){\circle*{1.5}}
\put(40,50){\vector(0,-1){30}}
\put(40,50){\circle*{1.5}}
\put(45,45){\vector(0,-1){25}}
\put(45,45){\circle*{1.5}}
\put(50,40){\vector(0,-1){30}}
\put(50,40){\circle*{1}}
\put(50,20){\circle*{1}}
\put(60,37.5){\makebox(0,0)[cc]{\ldots}}
\put(50,5){\makebox(0,0)[cc]{\ldots}}
\put(15,30){\circle{2.5}}
\put(15,27){\makebox(0,0)[cc]{$\op\theta_1$}}
\put(20,30){\circle{2.5}}
\put(20,27){\makebox(0,0)[cc]{$\op\theta_2$}}
\put(25,30){\circle{2.5}}
\put(25,27){\makebox(0,0)[cc]{$\op\theta_3$}}
\put(35,20){\circle{2.5}}
\put(35,17){\makebox(0,0)[cc]{$\op\theta_1$}}
\put(40,20){\circle{2.5}}
\put(40,17){\makebox(0,0)[cc]{$\op\theta_2$}}
\put(45,20){\circle{2.5}}
\put(45,17){\makebox(0,0)[cc]{$\op\theta_3$}}
\put(7.5,35){\dashbox{4}(75,45)[lt]{$\mathstrut$ Hybrid program ``bus''}}
\put(7.5,2.5){\dashbox{2}(75,30)[lb]{ Quantum data ``bus''}}
\end{picture}
\caption{Hybrid quantum processing unit.
 ($\op\theta_k \equiv \op\theta_{k,(p)} = e^{2 \pi i\, p \op\sigma_k}$)}
\label{fignet}
\end{center}
\end{figure}
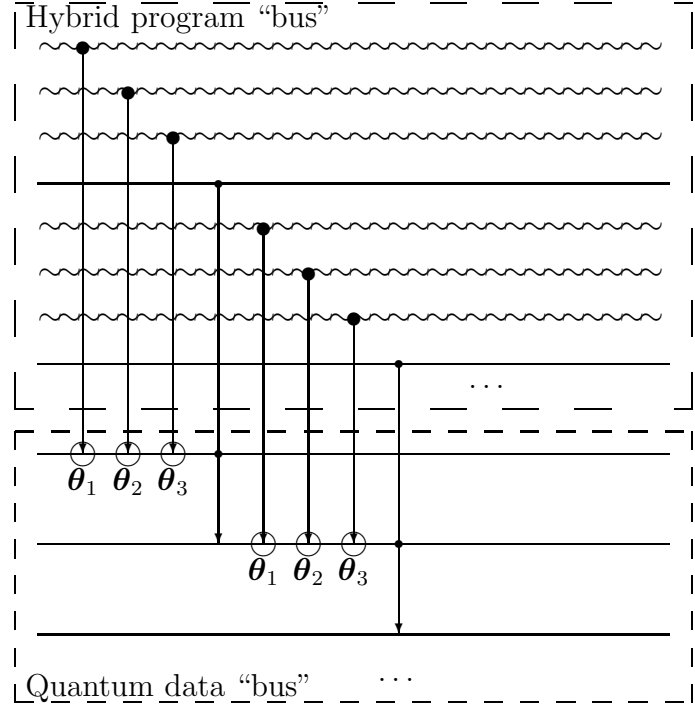

Here is also revisited \cite{Vla03} an alternative approach for strictly
universal quantum processor
--- to use continuous quantum variables in program register and discrete ones
for data, {\em i.e.\ hybrid} quantum computer \cite{Llo00}. In such a case
number of different programs is infinite and it provides possibility to
construct strictly universal hybrid quantum processor for initial
(``deterministic'') design described by \eq{qproc}. It is enough to
provide procedures for one-qubit rotations with three real parameters
together with some finite number of two-gates \cite{Cle99,Gate95}.

Generalization to hybrid system with program register described
by one continuous quantum variable and qubit data register is straightforward.
The states of program register may be described as Hilbert space of functions
on line $\psi(x)$. In coordinate representation a basis is
\begin{equation}
 \ket{q} = \delta(x-q); \quad \brkt{q}{\psi(x)} = \psi(q).
\label{q}
\end{equation}

Let's represent some continuous family of gates $\op u_{(q)}$ acting on data
state, say controlled qubit rotations
\begin{equation}
 \op\theta_{j,(q)}=\exp(2 \pi i q \op\sigma_j),
\label{thetaq}
\end{equation}
where $\op\sigma_j$ are three Pauli matrices.
It is possible
to write continuous analog of \eq{conddyn}:
\begin{eqnarray}
 &\QPU = \int dq \bigl(\ket{q}\bra{q} \ox \op u_{(q)}\bigr) , \\
 &\QPU\bigl(\psi(x)\ket{s}\bigr) = \int \delta(x-q) \psi(q) \ket{\op u_{(q)}s} dq=
 \psi(x)\ket{\op u_{(x)}s},
\end{eqnarray}
where $\ox$ is omitted because $\ket{\psi}\ket{s}$
can be considered as product of scalar function $\psi(x)$ on complex vector
$\ket{s}$.
Finally:
\begin{equation}
 \QPU (\ket{q}\ket{s}) = \ket{q}\ket{\op u_{(q)}s}.
\label{ctrQ}
\end{equation}

It is convenient also to use momentum basis, {\em i.e.}:
\begin{equation}
 \ket{\tilde p} = e^{i p x}; \quad
 \brkt{\tilde p}{\psi(x)} = {\textstyle\int} e^{-i p x} \psi(x) dx
 \equiv \tilde\psi(p).
\end{equation}
(where $\tilde\psi$ is Fourier transform of $\psi$)
and operator $\tilde\QPU$:
\begin{equation}
 \tilde\QPU = \int dp \bigl(\vxv{\tilde p} \ox \op u_{(p)}\bigr),
\end{equation}
\begin{equation}
 \tilde\QPU (\ket{\tilde p}\ket{s}) = \ket{\tilde p}\ket{\op u_{(p)}s}.
\label{ctrP}
\end{equation}

It has simpler physical interpretation. Let us consider scattering of some
scalar particle on quantum system with two states (qubit). Then
\eq{ctrP} can be written symbolically as:
\[
 {\rightwave^{\ket{\exp(i k x)}_1}}\ %
 {\mathop{{\bullet}\mkern-13mu{\nearrow}}\limits^{\ \ket{s}_2}}
 \quad \Longrightarrow \quad
 {\mathop{{\bullet}\mkern-13.5mu{\nwarrow}}\limits^{\op u_{(k)}\ket{s}_2}}\ %
 {\rightwave^{\ket{\exp(i k x)}_1}}
\]

Using such approach with hybrid program register (few continuous variables
for different qubit rotations and discrete ones for two-gates like CNOT), it
is possible to suggest design of universal quantum processor with qubits data
register.

Hilbert space of hybrid system with $k$ continuous and $M=2^m$ discrete
quantum variables can be considered as space of $\C^M$--valued functions
with $k$ variables
$$
 F(x_1,\dots,x_k) \colon \R^k \to \C^M.
$$

For construction of universal processor it is possible to use three
continuous variables for each
qubit together with discrete variables for control of two-qubit gates
(see Fig.~\ref{fignet}).

\section{Comparison with more traditional\\ models}

On the one hand, model of von Neumann quantum processor discussed in
present paper looks rather difficult for realization and author did
not hear about any proposal for experimental implementation of such
model. Of course, formally even {\sf Controlled-NOT} or more general
{\sf Controlled-U} quantum gates satisfy QPU design \eq{qproc}, but
they are too simple gates for serious discussions.

On the other hand, it is clear, that modern situation, when for
control of quantum processing are used some macroscopic and even
huge equipment may be not very convenient from point of view of
possible further plan to design {\em elementary quantum processors}.

There is yet another advantage of discussed approach. Paradigm of
hybrid quantum processor discussed above formally {\em includes even
more traditional design}, because continuous quantum variables without
superposition used in such model are analogues of classical variables
like different kind of electro-magnetic (including light) signals used for
construction of Hamiltonians for control of quantum gates almost in any
approach to quantum processing.

\end{document}